\documentclass[a4paper]{jpconf}
\usepackage{graphicx}
\begin{document}
\title{Probing New Physics with Astrophysical Neutrinos}

\author{Nicole F. Bell}

\address{School of Physics, The University of Melbourne, 
Victoria 3010, Australia}

\ead{n.bell@unimelb.edu.au}

\begin{abstract}
  We review the prospects for probing new physics with neutrino
  astrophysics.  High energy neutrinos provide an important means of
  accessing physics beyond the electroweak scale. Neutrinos have a
  number of advantages over conventional astronomy and, in particular,
  carry information encoded in their flavor degree of freedom which
  could reveal a variety of exotic neutrino properties.  We also
  outline ways in which neutrino astrophysics can be used to constrain
  dark matter properties, and explain how neutrino-based limits lead
  to a strong general bound on the dark matter total annihilation
  cross-section.
 
\vspace{2mm}
\noindent{\it Talk given at Neutrino 2008, Christchurch, New
    Zealand, 26-31 May 2008.}
\end{abstract}

\vspace{-4mm}

\section{Introduction}

Neutrino astronomy is in its infancy.  To date, the only neutrinos we
have observed from beyond our solar system are those from SN1987A.
Together with solar neutrinos, and those produced by cosmic ray
interactions in the atmosphere, these form the complete inventory of
astrophysical neutrinos that have been detected.  For distant
astrophysical objects, we currently have only upper limits on the
neutrino fluxes.  However, a plethora of exciting new experiments are
now coming on line with excellent prospects of detecting a signal.
The eagerly awaited era of neutrino astronomy is likely to prove
extremely revealing, both in terms of the properties of astrophysical
neutrino sources, and the properties of neutrinos themselves.  In this
article I concentrate on the later.

Studying the astrophysical neutrino flux produced by sources beyond the
solar system, may eventually be as revealing as the solar neutrino
flux has proven to be.  From an astrophysics point of view, neutrinos
have the advantage that (unlike cosmic ray protons) they are not
deflected by magnetic fields and thus their arrival direction points
back to the source.   In addition, they are not attenuated by
intervening matter. Neutrino astrophysics will thus allow us to see
further in the cosmos and deeper into astrophysical sources.  In
addition, the flavor composition of astrophysical neutrino fluxes may
encode important information about neutrino properties.

There are many interesting sources of high energy astrophysical
neutrinos, including cosmic accelerators such as gamma ray bursts,
supernovae remnants or active galactic nuclei.  Interactions of
accelerated nucleons in the vicinity of these sources lead to the
production of charged pions, and hence neutrinos via their decays.  If
the sources are optically thin, the neutrino fluxes may be related to
the fluxes of cosmic rays and gamma rays~\cite{Waxman,mpr}, while for
optically thick sources these constraints do not apply~\cite{mpr}.
There may even be  ``hidden sources'' for which the density of matter
is such that only neutrinos escape; see, for example,
Ref.~\cite{hidden,Mena}.  In addition, ``cosmogenic'' neutrinos are
generated via the interaction of high energy cosmic rays with the
cosmic microwave background.  Finally, dark matter annihilation or
decay may contribute a source of high energy neutrinos that are
detected in neutrino telescope experiments.

\section{Above the electroweak scale}

One of the most exciting prospects of neutrino astronomy is ability to
access physics beyond the electroweak scale.  For neutrinos with PeV
energies, the center of mass energy in a neutrino-nucleon interaction
is at the TeV scale.  At such high energies, the neutrino-nucleon
cross sections have not been measured and must be extrapolated from
lower energy data~\cite{Gandhi,crosssection1,crosssection2}.
Cross-sections either smaller or larger than the standard model
extrapolation could signal new physics contributions.  Possible effect
that could enhance neutrino cross-sections include the exchange of
towers of Kaluza-Klein gravitons~\cite{KK1,KK2} or the production of
black holes~\cite{BH1,BH2,BH3}.

Event rates in neutrino telescopes obviously depend on both the
neutrino fluxes and cross-sections.  However, it is possible to
disentangle flux and cross-section, since event rates in the up-going,
down-going, and earth-skimming directions have a different dependence
on neutrino cross-sections, due to absorption of neutrinos which
traverse the Earth~\cite{kusenkoweiler,BH1}.  Current Amanda data
place weak flux and cross-section constraints at center of mass
energies of order $\sim 1$ TeV~\cite{anchordoqui}, while IceCube and
other experiments have potential to make a measurement of these
parameters.

Many example of physics beyond the Standard Model may also show up in
neutrino telescopes.  For instance, in some supersymmetric models a
very distinctive process would be the production of long-lived NLSP
pairs, for which the signature in IceCube would be a pair of two
parallel charged tracks~\cite{Albuquerque:2003mi,ando,Ahlers:2007js}.

\section{Flavor Composition}

Neutrinos from astrophysical sources are expected to arise dominantly
from the decays of pions, which result in initial flavor ratios of
$\phi_{\nu_e}:\phi_{\nu_{\mu}}:\phi_{\nu_{\tau}} = 1:2:0$.  The fluxes
of each mass eigenstate are given by $\phi_{\nu_i}= \sum_{\alpha}
\phi_{\nu_\alpha}^{\rm source} |U_{\alpha i}|^2$, where $U_{\alpha i}$
are elements of the neutrino mixing matrix.  If we assume exact
$\nu_\mu$--$\nu_\tau$ symmetry ($\theta_{23}=45^\circ$ and
$\theta_{13}=0$) this implies that neutrinos are produced in the
ratios $\phi_{\nu_1}:\phi_{\nu_2}:\phi_{\nu_3} = 1:1:1$ in the mass
eigenstate basis, independent of the solar mixing angle.  Oscillations
do not change these ratios, only the relative phases between mass
eigenstates, which will be washed out by uncertainties in the energy
or distance since $\delta m^2 \times L/E \gg 1$.  An incoherent
mixture with the ratios $1:1:1$ in the mass basis implies an equal
mixture in any basis (since ${\cal U} I {\cal U}^{\dagger} \equiv I$)
and in particular in the flavor basis in which the neutrinos are
detected~\cite{111a,111b}.

Variation from the assumed $\nu_\mu$--$\nu_\tau$ symmetry lead to only
small deviations of the flavor ratios.  However, such deviations could
be used to measure neutrino mixing parameters, if sufficiently high
precision measurements of the astrophysical flux were to be
made~\cite{mixinganglesa,mixinganglesb,mixinganglesc}.  On the other
hand, the flavor composition of astrophysical neutrino fluxes may
encode important information about exotic neutrino properties.
Variations to the expected flavor ratios may reveal new physics such
as neutrino decay~\cite{astrodecay}, CPT violation~\cite{CPT},
oscillation to sterile neutrinos~\cite{pseudodirac,sterile1,sterile2},
and various other exotic scenarios~\cite{pas,enqvist,zhou}.

Neutrino decay can result in particularly large deviations to the
expected flavor ratios~\cite{astrodecay}.  For non-radiative decays
such as $\nu_i \rightarrow \nu_j + X$ and $\nu_i \rightarrow
\bar{\nu}_j + X$, where $X$ is a massless particle (e.g. a Majoron)
existing limits are quite weak.  If neutrinos are unstable, the cosmic
neutrinos detected may all be in the lightest mass eigenstate.  The
flavor composition of this lightest eigenstate is
$\phi_{\nu_e}:\phi_{\nu_{\mu}}:\phi_{\nu_{\tau}} = 5:1:1$ in the case
of the normal hierarchy, and $0:1:1$ in the case of the inverted
hierarchy.  For either hierarchy, this represents a large and
distinctive deviation to the expected flavor equality.

\begin{figure}
\begin{center}
\includegraphics[width=4.5in]{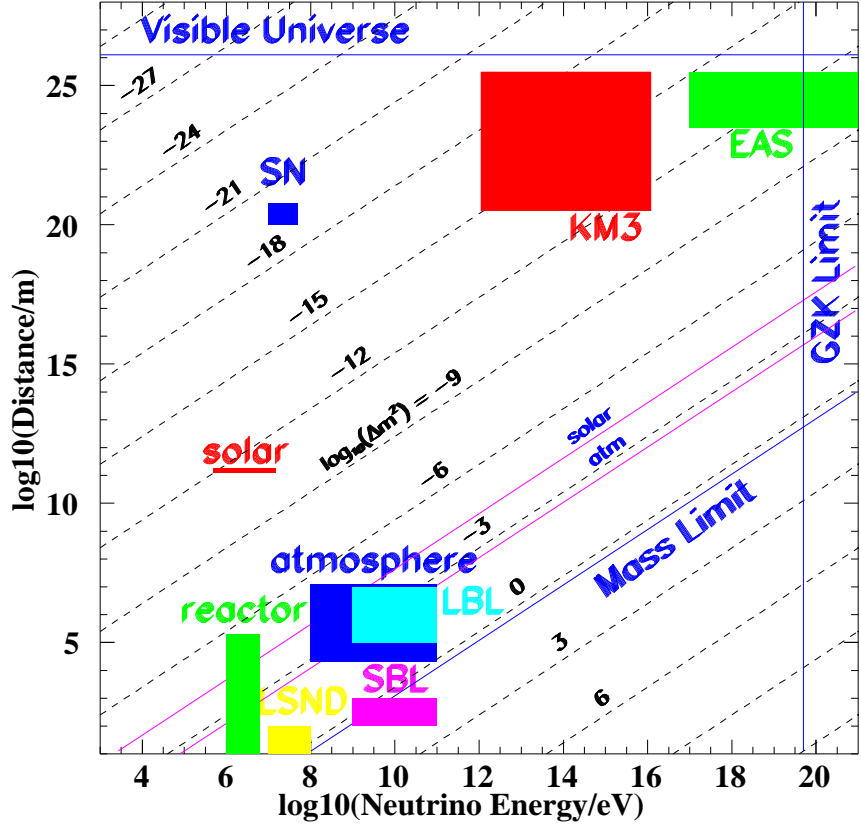}
\caption{ 
  The energy and distance ranges covered in various neutrino
  experiments.  The diagonal lines indicate the mass-squared
  differences (in eV$^2$) that can be probed with vacuum oscillations;
  at a given $L/E$, larger $\delta m^2$ values can be probed by
  averaged oscillations.  The shaded regions display the sensitivity
  of solar, atmospheric, reactor, supernova (SN), short-baseline
  (SBL), long-baseline (LBL), LSND, and extensive air shower (EAS)
  experiments.  The KM3 region describes the parameter space that
  would be accessible to a 1-km$^3$ scale neutrino telescope, given
  sufficient flux.  Figure taken from Ref.~\cite{pseudodirac}.}
  \label{fig:learned3}
\end{center}
\end{figure}

Another feature of astrophysical neutrino experiments is the enormous
distance scales at our disposal.  With neutrino from distant
astrophysical sources, we may do oscillation experiments with
baselines comparable to the size of observable Universe.  Given a
neutrino oscillation length scale of $\sim 2E/\delta m^2$,
cosmological scale baselines provide sensitivity to oscillations with
extremely small mass splittings~\cite{pseudodirac,sterile1,sterile2}.
In Fig.~\ref{fig:learned3} we show the $\delta m^2$ sensitivity of
various neutrino experiments.
An example in which such tiny mass splittings occur is the case of
pseudo-Dirac neutrinos, in which a Dirac neutrino is split into a pair
of almost degenerate Majorana neutrinos by the presence of tiny,
sub-dominant, Majorana mass terms.  In this scenario the active
neutrinos are each maximally mixed with a sterile partner with very
tiny $\delta m^2$.  The deviations to the astrophysical neutrino
flavor ratios due to oscillations driven by these tiny mass splittings
would be milder than those predicted for neutrino decay.  However, it
is a potential probe of tiny Majorana mass terms (and thus lepton
number violation) not discernible via any other means.

If neutrinos are produced via some mechanism other than conventional
pion decay, there will also be departures from the canonical flavor
ratios $1:1:1$.  One scenario is that where neutrons, produced in the
Galaxy by  photo-disintegration of heavy nuclei, decay to a pure
$\bar\nu_e$ flux~\cite{neutron1,neutron2}.  After oscillations wash
out phase information, this flux is transformed to the ratios
$\phi_{\nu_e}:\phi_{\nu_\mu}:\phi_{\nu_\tau} = 5:2:2$.  Another
possibility is a muon damped source in which charged pions decay to
muons and neutrinos, but the muon daughters loose energy before
decaying further~\cite{Rachen}.  The pure $\nu_\mu$ flux produced is
transformed by oscillations to
$\phi_{\nu_e}:\phi_{\nu_\mu}:\phi_{\nu_\tau} = 1:2:2$.

The flavor degree of freedom clearly carries important information
about both the astrophysical neutrino production mechanism, and exotic
physics in the neutrino sector.  An important question is whether a
given flavor signature is unique to a particular scenario.  However, a
number of the scenarios discussed above have large and distinctive
effect on the flavor ratios.  For example, it is difficult to see how
the neutrino decay of prediction  $\phi_{\nu_e}:\phi_{\nu_\mu} = 5:1$
could be replicated by another mechanism.

Neutrino flavor ratios will not be directly measured at neutrino
telescope experiments, but can be inferred from the ratios of
different types of events.  In an experiment like IceCube, the ratio
of muon tracks to shower events is likely to be most useful, and would
permit the $\nu_e$ fraction of the flux to be calculated.  In
Ref.~\cite{flavor}, it was found that a $\nu_e$ fraction of 1/3 (the
default prediction) could be measured to a range of approximately
0.2--0.4, provided the neutrino spectrum was also measured.

Double-bang and lollipop events, which are unique to $\nu_\tau$, would
provide important direct information on the size of the tau neutrino
flux~\cite{111a,tau}.  A double-bang event consists of a shower
initiated by a charged current interaction of $\nu_\tau$, followed by
a second shower initiated by the decay of the resulting tau
lepton. (Lollipop events consists of the second of these two showers,
along with a reconstructed tau lepton track.)  The detection threshold
for these $\nu_\tau$ events is a few PeV, and thus expected events
rates will be small.  However, given that the some exotic physics
scenario can lead to large deviations from the expected flavor
equality, even small numbers of events can provide important
information.

\section{Dark matter annihilation to neutrinos}

Dark matter (DM) may well be a source of high energy neutrinos that are
detected in neutrino telescope experiments, and there are a number of
techniques that use neutrinos to constrain dark matter cross sections.

WIMPS captured by the gravitational field of the Sun (and also the
Earth) accumulate in the center of the body and annihilate with a rate
proportional to the square of the DM number density~\cite{Press}.  All
products of such DM annihilation would be absorbed in the Sun, with
the exception of neutrinos.  Therefore, since neutrinos produced via
solar fusion processes have typical energies of $\sim 10$ MeV, an
observation of neutrinos with GeV energies or above emanating from the
solar core would be strong evidence for dark matter.  Such techniques
enable Super-Kamiokande~\cite{Desai}, Amanda~\cite{Ackermann} and
other experiments to place competitive constraints on the WIMP-nucleon
scattering cross-section.

High energy neutrinos may also be produced via DM annihilation or
decay in galactic halos.  In this case, we look for fluxes produced in
the Milky Way, in other galaxies, or for a diffuse flux arising from
annihilation or decay in all halos throughout the Universe.
Neutrinos, despite being generally harder to detect than, e.g., gamma
rays, in fact provide important information and surprisingly strong
bounds on the {\it total} dark matter annihilation rate~\cite{DMnu}.

Let us assume that the DM is the lightest particle beyond those in the
Standard Model (SM).  It then follows that all dark matter
annihilation products must be Standard Model particles, as any other
(new) particles must be kinematically inaccessible.  Among SM final
states, it is clear that all but neutrinos will inevitably produce
gamma rays.  Quarks and gluons hadronize, producing pions, where
$\pi^0 \rightarrow \gamma \gamma$, and the decays of weak bosons and
tau leptons also produce $\pi^0$.  The stable final state $e^+ e^-$ is
not invisible, since it produces gamma rays either through
electromagnetic radiative corrections or energy loss processes, while
the final state $\mu^+ \mu^-$ produces $e^+ e^-$ via decays.  Given
that limits on the $\bar{\nu} \nu$ final state will be weaker than the
limits on all other final states, we can set a conservative upper
limit on the {\it total} annihilation cross-section by setting the
branching ratio to the $\bar{\nu} \nu $ final state at $Br(\bar{\nu}
\nu) = 100\%$.

\begin{figure}
\begin{center}
\includegraphics[width=4.5in]{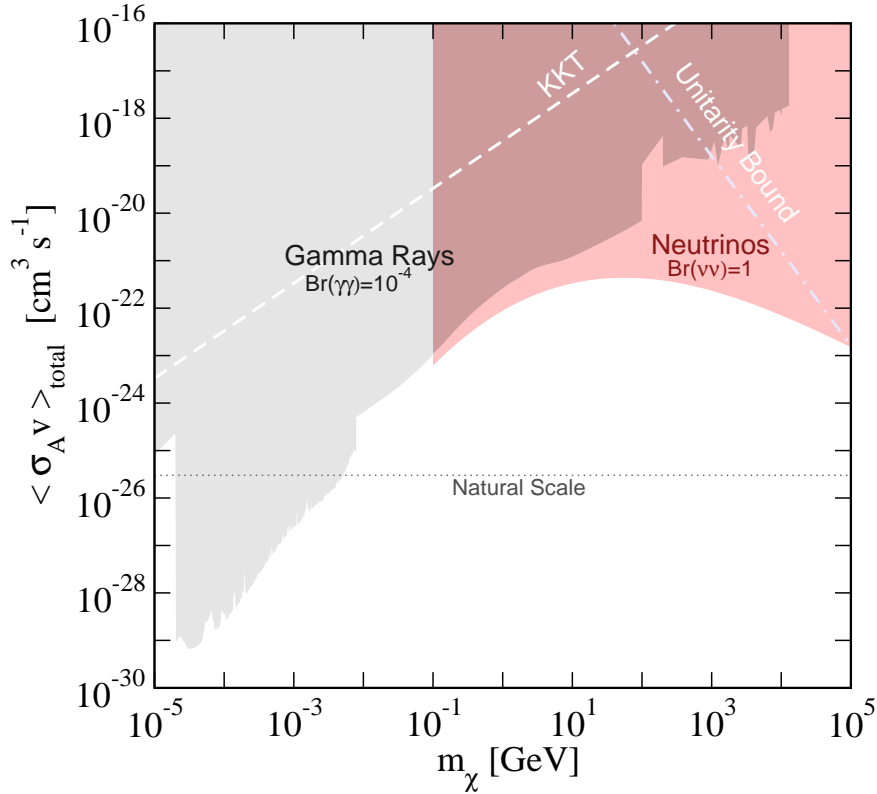}
\caption{ Neutrino and gamma-ray limits on the dark matter total
  annihilation cross section in galaxy halos, selecting
  $Br(\gamma\gamma) = 10^{-4}$ as a conservative value.  The general
  unitarity bound is shown for comparison, while the KKT limit denotes
  the point at which annihilations would significantly modify dark
  matter halo density profiles~\cite{KKT}.  The overall bound on the
  total cross section at a given mass is determined by the strongest
  of the various upper limits.  Figure taken from Ref.~\cite{Mack}.}
  \label{fig:sigmagg-fig3}
\end{center}
\end{figure}

The most straightforward approach to bound the $\textrm{\small{DM}} +
\textrm{\small{DM}} \rightarrow \bar{\nu} \nu$ cross section is to use
the cosmic diffuse neutrino flux arising from dark matter annihilation
in all halos throughout the Universe~\cite{DMnu}.  The neutrino
annihilation rate depends on the average of $n_{\textrm{\tiny{DM}}}^2$
(where $n_{\textrm{\tiny{DM}}}$ is the DM number density) which is
enhanced by the clustering of dark matter in halos, while the
monochromatic neutrino energy is smeared by redshift to form a broader
spectrum.  A complementary approach, with comparable or slightly
better sensitivity, is to consider the signal from annihilations
within our Galactic halo~\cite{Yuksel}.
In order for the annihilation flux to be detectable, it must be larger
than the atmospheric neutrino background.  We may adopt the
conservative criteria that the signal be as large as the angle
averaged atmospheric neutrino background, and use the Super-Kamiokande
atmospheric neutrino measurements~\cite{superkatm} to bound the
possible neutrino flux arising from dark matter annihilation.

In Fig.~\ref{fig:sigmagg-fig3} , we show the constraints on the {\it
  total} dark matter annihilation cross-section, obtained by
conservatively setting the branching ratio to neutrinos at 100\% (the
figure displays the Milky Way constraints derived in
Ref~\cite{Yuksel}).  Also shown are constrains on the annihilation
cross-section obtained by assuming a $10^{-4}$ branching ratio to the
state $\gamma\gamma$~\cite{Mack}.  The neutrino results are
surprisingly strong, particularly for large dark matter mass.  In
particular, they are more stringent than the general unitarity
bound~\cite{Hui} over a large range of masses, and strongly rule out
proposals in which annihilation rates are large enough to modify dark
matter halos (denoted by KKT~\cite{KKT} in
Fig.~\ref{fig:sigmagg-fig3}).

The technique to constrain the dark matter total annihilation
cross-section can be applied to MeV energies using the
Super-Kamiokande data~\cite{sergioMeV}, and analogous bounds on the DM
decay rate can also be derived~\cite{sergiodecay}.  Note that although
we have set the branching ratio to neutrinos at 100\% (in order to
derive a conservative and model independent bound) final state
neutrinos will inevitably by accompanied by gamma rays due to
electroweak radiative corrections.  However, these gamma ray
constraints on the annihilation cross-section are weaker than or
comparable to those obtained directly with
neutrinos~\cite{Kachelriess,Bell,Dent}.

\end{document}